\documentclass{osa-article}

\journal{oe}
\articletype{Research Article}
\usepackage{multirow}
\usepackage{multicol}
\usepackage{algorithm}
\usepackage{algorithmic}
\usepackage[export]{adjustbox}
\usepackage[utf8]{inputenc}

\def\R{\mathbb{R}}          								
\newcommand\bth[0]{{\boldsymbol{\theta}}} 

\newcommand{\bst}[0]{\boldsymbol{t}}

\newcommand\bsu[0]{{\boldsymbol{u}}}
\newcommand\bsv[0]{{\boldsymbol{v}}}
\begin{document}

\title{Mechanical Artifacts in Optical Projection Tomography: Classification and Automatic Calibration}

\author{Yan Liu, \authormark{1} Jonathan Dong,\authormark{1} Thanh-an Pham,\authormark{1} Fran\c cois Marelli, \authormark{2,3} and Michael Unser\authormark{1, *}}

\address{\authormark{1} Biomedical Imaging Group, \'Ecole polytechnique f\'ed\'erale de Lausanne, Station 17, 1015 Lausanne, Switzerland\\
\authormark{2} Idiap Research Institute, 1920 Martigny, Switzerland\\
\authormark{3} \'Ecole polytechnique f\'ed\'erale de Lausanne, 1015 Lausanne, Switzerland\\
}

\email{\authormark{*}michael.unser@epfl.ch} %% email address is required

% \homepage{http:...} %% author's URL, if desired

%%%%%%%%%%%%%%%%%%% abstract %%%%%%%%%%%%%%%%
%% [use \begin{abstract*}...\end{abstract*} if exempt from copyright]

\begin{abstract}
Optical projection tomography (OPT) is a powerful tool for biomedical studies. 
It achieves 3D visualization of mesoscopic biological samples with high spatial resolution using conventional tomographic-reconstruction algorithms. 
However, various artifacts degrade the quality of the reconstructed images due to experimental imperfections in the OPT instruments. 
While many efforts have been made to characterize and correct for these artifacts, they focus on one specific type of artifacts, whereas a comprehensive catalog of all sorts of mechanical artifacts does not currently exist.
In this work, we systematically document many mechanical artifacts.
We rely on a 3D description of the imaging system that uses a set of angular and translational parameters. 
We provide a catalog of artifacts.
It lists their cause, resulting effects, and existing correction methods. 
Then, we introduce an automatic calibration algorithm that is able to recover the unknown system parameters fed into the final 3D iterative reconstruction algorithm for a distortion-free volumetric image. 
Simulations with beads data and experimental results on a fluorescent textile fiber confirm that our algorithm successfully removes miscalibration artifacts in the reconstruction. 
\end{abstract}
%\newpage
%%%%%%%%%%%%%%%%%%%%%%%%%%  body  %%%%%%%%%%%%%%%%%%%%%%%%%%
\section{Introduction}
Conventional optical microscopy such as confocal microscopy is limited to the imaging of relatively thin samples.
This limitation can be partially overcome with optical projection tomography (OPT) which was invented by Sharpe in 2002 \cite{Sharpe2004}.
Over the years, OPT has become a mature tool for the  production of high-resolution 3D images of biological samples at mesoscopic scale \cite{Lindsey2017 ,Schmidt:21} in a brightfield \cite{Wang:07, bassi:11} or fluorescence \cite{Walls2007, Walls2005, McGinty:11, Torres2021} configuration. 
OPT is widely used in a variety of applications, such as the mapping of the distribution of proteins in embryos \cite{Sharpe2002, Sharpe2004}, the localization of  metastases in lymph nodes \cite{Torres2021}, the display of vascular networks and amyloid depositions in the mouse brain model to study Alzheimer's disease \cite{Lindsey2017}, and the imaging of the spatial arrangement of intestinal villi \cite{Schmidt:21}. 

Since OPT is a tomographic-imaging technology, it falls into the same category as X-ray computed tomography (CT), single-photon electron computed tomography and electron tomography \cite{Sharpe2004}. 
The contrast mechanism that these technologies rely upon is either the attenuation or the emission function of rays of light. 
For OPT, the rays follow optical straight lines that are geometrically only approximately straight. They project the 3D inner structure of the sample onto a 2D detector plane.
The mathematical tool to describe such a straight-ray projection is the Radon transform \cite{Natterer}.
The associated inverse problem is to reconstruct the 3D volume from the set of 2D projections acquired at various spatial positions of the sample. 
The discrete inversion formula of the Radon transform, the filtered backprojection (FBP) algorithm, is efficiently implemented and widely used in practice \cite{Torres2021, Schmidt:21, Vinegoni:09}.  

While OPT achieves high-resolution 3D imaging at a relatively low cost \cite{Ramirez:19, Zhang:20}, the reconstructions often suffer from artifacts that impact the quality of the images due to several types of model mismatch. 
Among them, mechanical errors in the imaging system play a non-negligible role, especially in low-cost OPT systems \cite{Ramirez:19}.
The most common type of mechanical error is an offset of the rotation axis which results in a misaligned center of rotation during the experiment.
This typically leads to distortions including double edges \cite{Walls2005} or circles \cite{Tang:16} in the reconstruction, depending on the type of sample.
Many 2D methods have been proposed to address this issue, including maximum variance of a reconstructed slice under a set of guesses of the true rotation \cite{Walls2005, Torres2021}, sinogram unification of both fluorescent and bright-field OPT \cite{Tang:16}, center of mass or image registration \cite{Donath:06}, and total-variation regularization \cite{michalek_2015}. 
These 2D methods assume that the tilt of the rotation axis is negligible.
Their extension to small tilt angles introduces a height dependence of the rotation axis \cite{Ramirez:19, Torres2021}. For larger tilt angles, a method to account for tilt in 3D is still missing. 
Another type of mechanical error that has not been well studied is the angular errors of the rotation motor, which will result in seagull-shaped artifacts for point-like objects, as we show in Section 2. 
Artifacts may also arise due to optical effects. For example, the assumption that optical straight paths coincide with geometric ones need to be abandoned, thus requiring variations of the conventional Radon transform \cite{Walls2007, Koskela19, Koljonen19, Trull2017}. 
Other types of model mismatch such as mismatches in the refractive index \cite{Birk2011, Antonopoulos:14, liu2022artifacts}, illumination fluctuations \cite{Walls2005}, spatial or temporal variations of the sensitivity of the detectors, their linearity, and background noise \cite{Birk2011, Walls2005} can lead to various distortions as well, degrading the reconstruction quality. 

In this paper, we first strive to provide a comprehensive catalog of many types of mechanical artifacts as a reference for OPT practitioners to assist them to calibrate their experiments. 
We rely on point-like objects (a very popular tool in the characterization of OPT setups) to demonstrate the cause and resulting appearance of each artifact.
Then, we introduce a 3D auto-calibration algorithm to remove the artifacts listed in our catalog, for the convenience of OPT experimentalists. 

This paper is organized as follows. 
In Section 2, we first describe the imaging geometry of OPT using a set of system parameters (angles and shifts).
We point out typical places where the most critical mechanical errors occur.
Then, we present a catalog of artifacts along with a detailed description of the three most common types of mechanical errors. 
In Section 3, we introduce a 3D forward model that characterizes all sorts of mechanical errors and a corresponding joint-reconstruction-calibration algorithm at coarse scale.
It is able to recover the unknown set of characterizing angles including the rotation and tilt angles. 
The refined system parameters are then fed into the 3D reconstruction algorithm to achieve an image without miscalibration artifacts at a finer scale.
Results on both simulated measurements of beads and experimental data of a fluorescent textile fiber are presented to validate our algorithms.

\section{Characterization of Mechanical Artifacts}
\begin{figure}[t]
\includegraphics[width=0.7\textwidth, center]{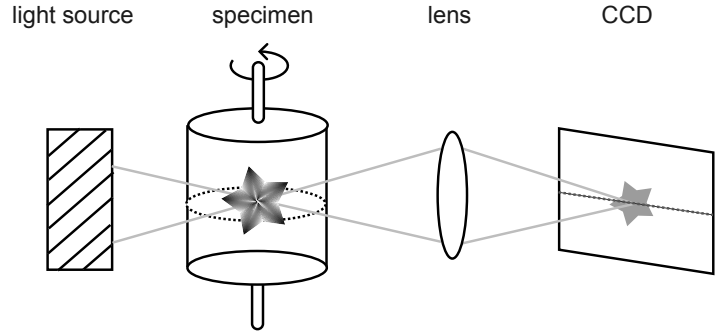}
\caption{Diagram of OPT. 
The light that passes through or is emitted by the sample gets imaged on a CCD camera using a lens system. 
In this ideal configuration, a slice of the sample perpendicular to the rotation axis corresponds to the information in the temporal sequence recorded by a row of pixels on the CCD camera.
}
\label{fig:setup}
\end{figure}
%--
\begin{figure}[t]
\includegraphics[width=0.5\textwidth, center]{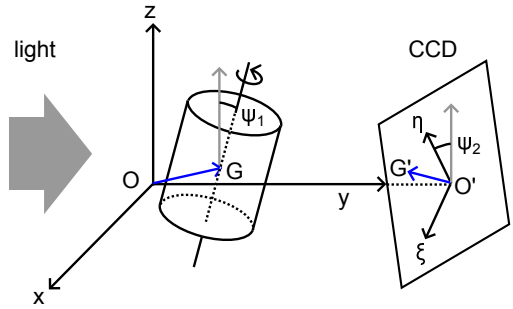}
\caption{
OPT imaging system with mechanical errors. 
(The lens system is omitted here for simplicity.)
The cylinder represents the tilted sample that rotates around its orientation axis.
The center of the cylinder is translated to $G$. 
The detector plane is described by a 2D coordinate system ($O'-\xi\eta$) perpendicular to the optical axis ($y$ axis). The projection of $O$ and $G$ in the detector plane are $O'$ and $G'$, respectively.
The angle $\psi_1$ represents the out-of-plane tilt angle between the orientation axis of the sample and the $z$-axis direction (gray vertical arrow).
The rotation angle $\psi_2$ represents the in-plane tilt angle of the detector plane around the optical axis.
}
\label{fig:geometry}
\end{figure}
%%%%%%%%%%%%%%%%%%%%%%%%%%%%%%%%%%%%%%%%%%%%%%%%%%%

\subsection{Common OPT Geometry and Reconstruction Algorithms}

The general OPT geometry is illustrated in Fig. \ref{fig:setup}. A 3D sample is placed in a rotating cylinder and imaged on a camera. 
This camera records 2D projections of either the absorption of the sample (transmission OPT) \cite{Koskela19} or its emitted fluorescence (emission OPT) \cite{Walls2007, vanderHorst:16}.
These tomographic projections, not considering more intricate models of light propagation, are enabled in reason of the very low numerical aperture of the OPT imaging system, in sharp contrast with conventional optical microscopy.
The sample is rotated around the axis of the cylinder for tomographic reconstruction. 

We place the sample in a 3D coordinate system ($O-xyz$) (Fig. \ref{fig:geometry}). In an ideal setup \cite{Sharpe2002}, the center of the cylinder is at $O$ while the rotation axis would be perfectly parallel to the $z$-axis. 
In our refined model, the center of the sample cylinder is at $G$ and its orientation is described by the three angles $\bth=(\boldsymbol{\varphi},\psi_1,\psi_2)$ for a rigid-body rotation. 
The $P$-dimensional vector $\boldsymbol{\varphi} = (\varphi_1, \ldots, \varphi_P)$ represents the set of $P$ rotation angles around the rotation axis, one for each captured projection. 
The tilted rotation axis is described by the remaining two angles $\psi_1$ and $\psi_2$ which are scalars that represent the out-of-plane and in-plane tilt, respectively, common to all views.
The detector plane is described by a 2D coordinate system ($O'-\xi\eta$) that is perpendicular to the optical axis. 

Many reconstruction algorithms are available for OPT, from FBP \cite{Koljonen19, Darrell2008} to optimization-based methods \cite{Darrell_MLE, Trull2017}. 
They mainly come from the field of CT, a canonical tomography application using X-rays with a similar parallel-beam geometry \cite{CT2001}. 
In the simplistic model where the rotation axis is assumed to be aligned with the $z$-axis and when the in-plane tilt vanishes, one horizontal plane of the sample corresponds to one line of the camera. Many reconstruction algorithms thus operate on camera images line by line, reconstructing the object plane by plane. 

A mismatch between the geometry of the setup and that assumed by the reconstruction algorithm obviously impact the quality of the final reconstruction. This induces artifacts that have been reported previously in the literature, along with dedicated procedures to correct for them. 

\subsection{Previously Reported Artifacts and Correction Techniques}

It may occur that the center of rotation (COR) $G$ is off-center, which is a common issue for OPT experiments. 
The consequences depend on the direction of this translation mismatch. A constant shift in the $x$-axis direction (parallel to the detector plane) has been reported in both OPT \cite{Walls2005, Tang:16} and X-ray CT \cite{Donath:06}. 
It will result in a constant shift in the 2D projections generating ringing artifacts: each bright point in the sample will be imaged into a circle after OPT reconstruction. 
For example, Tang \textit{et al.} reported circles in the imaging of a zebrafish embryo \cite{Tang:16}. 
Due to the finite depth-of-focus in OPT, these circles may take different flavors. 
Walls \textit{et al.} observed a so-called double-edge artifact while imaging the cardiac region of a mouse embryo \cite{Walls2005} and Donath \textit{et al.} observed tuning-fork artifacts of a point object \cite{Donath:06}. 
A shift in the $y$ axis (the direction of light propagation) will result in out-of-focus blur artifacts due to limited depth-of-field that is unique to OPT \cite{Koskela19}.

Different methods are available to correct for an off-center COR in both OPT and the field of X-ray CT. 
Walls \textit{et al.} proposed a heuristic method that calculates the variance of each reconstruction of a 2D slice using a range of assumed shift values in a sinogram: the shift value that produces the maximum variance serves as a close guess for the true shift \cite{Walls2005}.
This method is fast to implement and yields satisfactory reconstructions when the volume is not too sparse; 
it is commonly used as a preprocessing step by OPT practitioners \cite{Schmidt:21}. 
The center-of-mass method is another popular method to determine the center position in OPT. 
It is based on the property that the center of mass of the object is projected onto the center of mass of the sinogram.
Hence, a shift of the object will result in a corresponding shift of the center of mass in the sinogram. 
This constant shift is found by solving a linear system \cite{Donath:06}. 
Other methods such as image registration \cite{Donath:06} and methods that take advantage of the symmetrical structure of certain samples \cite{Zhang:20} are also used in OPT but with limitations because they require a priori information. 

The tilt of the rotation axis in 3D can be described by a combination of the out-of-plane and in-plane tilt.
Only in-plane tilts of a small angle have been studied. 
With this assumption, each transverse slice of the sample still approximately corresponds to a row of pixels on the camera; 
then, the compensation of the 3D tilt reduces to the 2D problem of finding the true center of rotation for each slice. 
This is done using the same techniques as for dealing with an off-center COR. 
For example, Torres \textit{et al.} \cite{Torres2021} used the maximum-variance method \cite{Walls2005} to find the COR at two different heights.
They then performed a linear regression to obtain the COR shift at any intermediate horizontal plane. 

Previous techniques are dedicated to the correction of a single type of mechanical artifact. 
To the best of our knowledge, there has been no description of the effect of other tilt angles. 
Moreover, the tilt-correction strategy is limited to small tilt angles, and no robust correction method has been proposed yet.

\subsection{Catalog of Mechanical Artifacts}
%--
\begin{table}[t]
    \centering
    \begin{tabular}{c|c|c|c|c|c}
    \hline
    \hline
        \multicolumn{2}{c|}{\multirow{2}{*}{\textbf{Error}}} & \multicolumn{1}{c|}{\multirow{2}{*}{\textbf{Visual clues}}} & 
        \multicolumn{2}{c|}{\textbf{Dependence}} &
        \multirow{2}{*}{\textbf{Reference}} 
        \\ \cline{4-5}
        \multicolumn{2}{c|}{} & & \textbf{FOV} & $z$ &
        \\ \hline
        \multirow{3}{*}{Translation} &
        $x$ axis & Circle, double edge & \multirow{3}{*}{No} & \multirow{3}{*}{No} & \cite{Tang:16, Donath:06, Birk:10, Walls2005, Zhang:20}
        \\ \cline{2-3} \cline{6-6}
        & $y$ axis & Out-of-focus blur & & & \cite{vanderHorst:16, Ancora2017, Koskela19}
        \\ \cline{2-3} \cline{6-6}
        & $z$ axis & None & & & N/A
        \\ \hline
        \multirow{2}{*}{Tilt of rotation axis} &
        Out-of-plane $\psi_1$ & Various & \multirow{2}{*}{Yes} & \multirow{2}{*}{Yes} & N/A
        \\ \cline{2-3} \cline{6-6}
        & In-plane $\psi_2$ & Various & & &  \cite{Torres2021, Ramirez:19}
        \\ \hline
        Rotation angle & Accumulative & Seagull & Yes & No & N/A
        \\ \hline
        \hline
    \end{tabular}
\caption{Table of setup-related imaging artifacts and their dependence on the field of view (FOV) and on height (location along the $z$ axis).}
    \label{tab:dict}
\end{table}

\begin{figure}[tbh!]
\centering
\includegraphics[width=\linewidth]{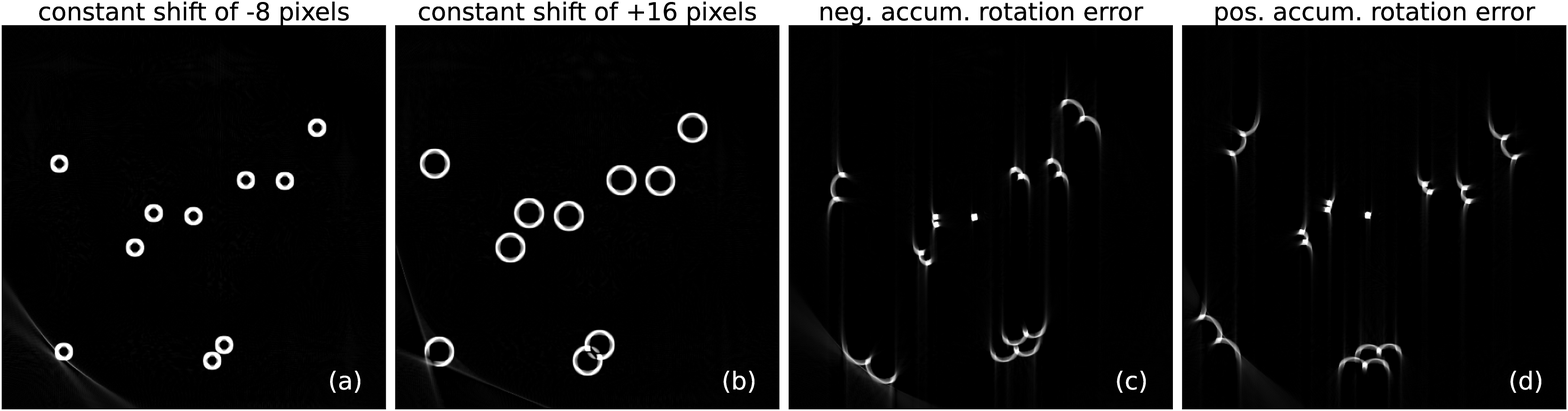}

\includegraphics[width=\linewidth]{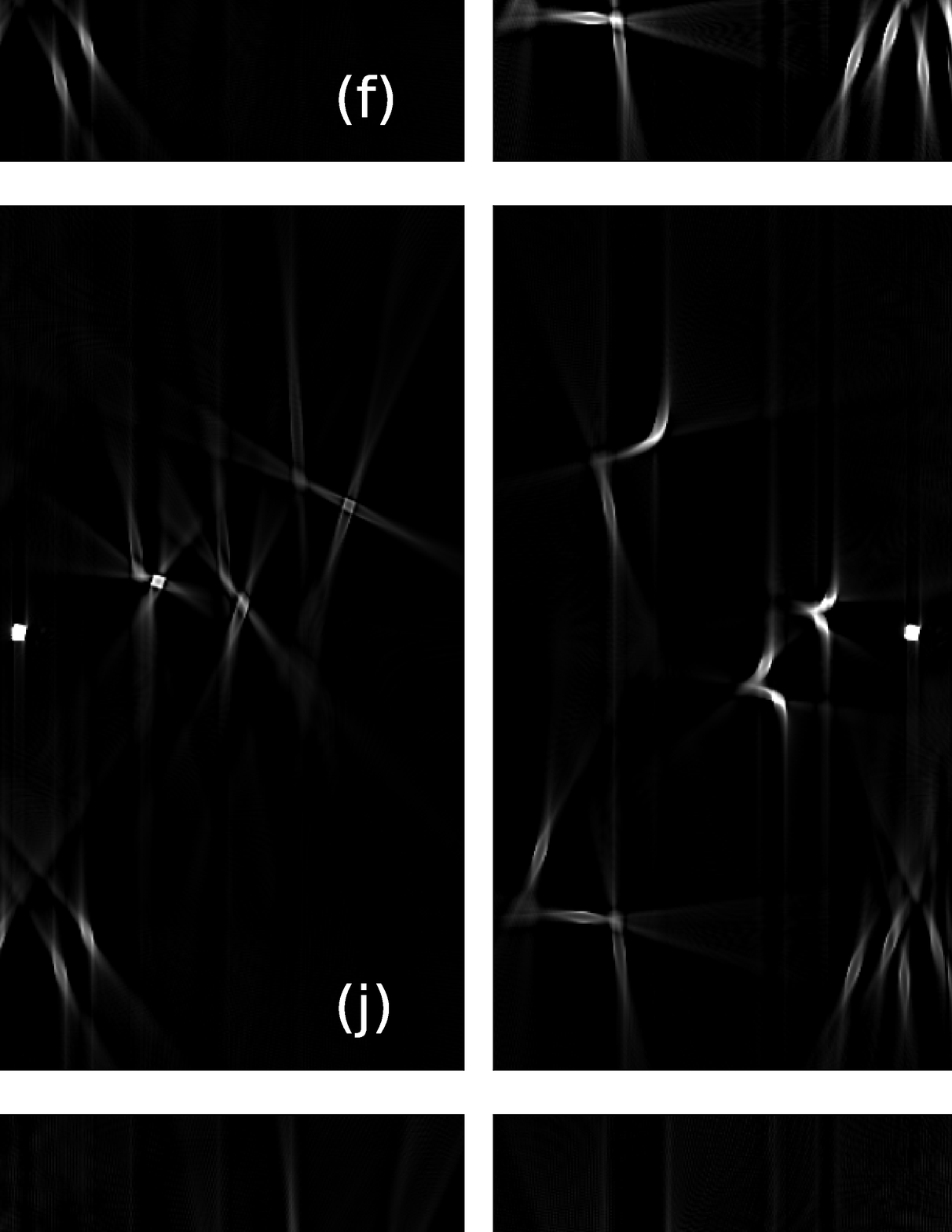}
\caption{Artifacts in the reconstruction of simulated beads due to a controlled mismatch the geometric parameters of our refined model and the simplistic geometry assumed by traditional reconstruction methods.
(a)-(b) Circle artifacts due to a constant shift of -4 pixels (a) and 8 pixels (b) of the COR in the $x$ direction. 
(c)-(d) Seagull artifacts due to a negative (c) and positive (d) random accumulative error with maximum amplitude 0.03 degrees in the rotation angles. 
(e)-(p) Various artifacts due to the tilt of the rotation axis.
(e)-(h) Reconstruction at slice 192 (64 slices below the central slice).
(i)-(l) Reconstruction at slice 256 (central slice in $z$-axis).
(m)-(p) Reconstruction at slice 384 (128 slices above the central slice).
Images are slightly saturated for better visualization.
}
\label{fig:artifacts}
\end{figure}

The geometry of OPT is described by the three-dimensional coordinates $G (x_G, y_G, z_G)$ of the center of the cylinder center  and the angles $\bth=(\boldsymbol{\varphi},\psi_1,\psi_2)$. 
In this parameterization, all parameters are fixed during an OPT acquisition except the rotation angles $\boldsymbol{\varphi}$. 
To characterize different mechanical imperfections, we introduce a catalog of the imaging artifacts that result from a mismatch between the ``true'' value and the value assumed at reconstruction for each of the parameters of out refined model. 
This catalog can be useful for experimentalists to identify artifacts they may encounter in the reconstructions, so as to correct for them either experimentally (re-calibration) or computationally. 
We summarize in Table \ref{tab:dict} the results with references to previous works when relevant, accompanied with visualizations shown in Fig. \ref{fig:artifacts} based on numerical simulations using point-like beads, a popular tool to characterize OPT systems.

The numerical simulation is done using Tomosipo, a convenient and versatile tomographic toolbox for 3D simulations and reconstructions of any geometric setup \cite{Hendriksen:21}.
We use a ground-truth object of 10 scattered beads of size $6^3$ pixels in the FOV with their centers on the same plane perpendicular to the $z$ axis. This object is placed in a cube of size $512^3$ pixels.
The number of rotation angles is 1200 over the range of 0 to 360 degrees to generate a set of 2D projections.
The location of this plane was moved along the $z$ axis to observe the dependence of a certain type of mechanical artifacts on the $z$ location.
We added different types of errors in the geometry of the forward model to simulate a realistic imperfect OPT system.
Then, we used the standard FBP algorithm which corresponds to a generic imaging geometry with no correction of mechanical errors. 

Shifts along the $x$-axis (orthogonal to the axis of rotation and light propagation) result in circle artifacts.
The size of artifacts is uniform across the FOV; however, their severity, as is indicated by the radius of the circle, depend on the absolute value of the shift. 
The larger it is, the bigger the circle artifacts will be (Fig. \ref{fig:artifacts}(a) and \ref{fig:artifacts}(b)). 
This characterization can be applied to any object, by considering it as a sum of point-like elements. For continuous samples, their borders will appear as double edges as reported in previous OPT experiments \cite{Walls2005}.
Shifts along the $y$-axis (direction of light propagation) cause an out-of-focus blur when parts of the sample move out of the depth-of-field of the imaging system. 
Shifts along the $z$ axis (rotation axis) cause an overall shift of the reconstruction in the $z$ direction, but do not generate artifacts as long as the object is still within the field of view of the camera. 

The tilt angles $\psi_1$ and $\psi_2$ generate artifacts of different shapes, especially when the mismatch is not small. 
To study the impact of the tilt of the rotation axis in different directions and levels of severity, as well as its dependence on the vertical location along $z$ axis, we moved the $z$ location of the beads to three different transverse slices: slice 256 (central slice), slice 192 (64 slices below the central slice), and slice 384 (128 slices above the central slice).
We added errors in the two tilt directions described previously with two values: 5 and 10 degrees. The result is shown in Fig. \ref{fig:artifacts}(e) - \ref{fig:artifacts}(p).

The first two columns of images of the second to the fourth rows in Fig. \ref{fig:artifacts} show that an out-of-plane tilt  results in various artifacts that appear more severe close to the boundary of the image. 
The next two columns of images of the second to the fourth rows in Fig. \ref{fig:artifacts} show that in-plane tilt in general leads to more severe artifacts of various shapes compared to the other direction. 
The dependence of the size of the artifacts on the location along the vertical direction is stronger than that of the tilt in the other direction.
The central slice shows the least artifacts in the third row of Fig. \ref{fig:artifacts}, while slices far away from it show bigger artifacts in the second and last row of Fig. \ref{fig:artifacts}.
As the tilt angle increases, the artifacts get stronger if we compare the first two columns or the last two columns of the tilt images in Fig. \ref{fig:artifacts}.
We notice the circle artifacts in Fig. \ref{fig:artifacts}(g), \ref{fig:artifacts}(h) and \ref{fig:artifacts}(o) which indicate that, when the impact of the tilt is small, it can be approximated by a shifted COR problem. 
In the case of a relatively large unknown tilt, such an approximation does not hold anymore, as seen in Fig. \ref{fig:artifacts}(p). 
To the best of our knowledge, correction for the tilt in 3D has not been studied yet.

The last parameter describing our refined OPT geometry is the set of rotation angles $\boldsymbol\varphi$. 
On one hand, since each rotation angle is different for each camera image, a random mismatch on this angle does not induce artifacts but introduces noise that degrades the image quality. 
On the other hand, a miscalibration of the rotation motor may introduce a positive or negative drift in the rotation angle, which accumulates over time. 
We simulated this kind of model mismatch and observed seagull-shaped artifacts in the reconstruction (Fig. \ref{fig:artifacts}(c) and \ref{fig:artifacts}(d)). 
Depending on the sign of the random accumulative error, the seagull is either ``flying'' toward or outward the center of the image. 
The size of the seagull in each transverse slice depends on the distance of the bead relative to the boundary of the object. 
The closer to the boundary, the bigger the seagulls will be while the bead located at the exact center of the slice appears unaffected since it coincides with the rotation axis (Fig. \ref{fig:artifacts}(c) and \ref{fig:artifacts}(d)). 

In Section 3, we introduce an automatic calibration algorithm to correct for the mechanical errors presented in Table \ref{tab:dict}. 
We start with a 3D forward model that fully characterizes our refined OPT geometry, including the 3D tilt that is out of reach of existing 2D methods.
Then, we formulate the calibration of the system parameters as a multiscale joint reconstruction-calibration optimization problem.
This multiscale scheme allows us to overcome the memory bottleneck of 3D models and helps us to accelerate the calibration. 
Once the system is calibrated, we are able to reconstruct an artifact-free volume.

\section{Automatic Calibration of Mechanical Artifacts}
In this section, we propose a computational framework that automatically calibrates the parameters of our refined model of an OPT imaging system. 
This framework is able to detect the model mismatch between the simplistic forward model and the measurement data.
It outputs a set of calibrated system parameters and allows us to improve the reconstruction of the 3D volume.
We first show how to characterize the three types of mechanical errors mentioned in Section 2.3. 
Then, we present our multiscale joint reconstruction-calibration algorithm and show how to remove the artifacts in both simulated and experimental data.
%=========================================================
\subsection{Forward Model and Characterization of Mechanical Errors}

The coordinate system to describe the OPT geometry is the same as described in Fig. \ref{fig:geometry}. When there exist mechanical errors in the system, we describe the angular errors as the perturbation vector $\boldsymbol{\delta}=(\boldsymbol{\delta}_{\boldsymbol{\varphi}},\delta_{\psi_1},\delta_{\psi_2})$.
The actual angles $\bth={\bth}^*+\boldsymbol{\delta}$ can be expressed as a sum of the error vector $\boldsymbol{\delta}$ and the ideal angle vector ${\bth}^* = (\boldsymbol{\varphi}^*, \psi_1^*, \psi_2^*)$ that characterizes a simplistic OPT system: equidistant rotation angles $\boldsymbol{\varphi}^*$ between 0 and 360° and $\psi_1^* = \psi_2^* = 0$.
In addition, translation error of the sample is described by a 2D vector $\bst=(t_1,t_2)\in\R^2$ in the detector plane. 

We omit optical effects and complex PSF models to keep the forward model computationally efficient, as is done in most OPT experiments. 
This omission is often acepted in the focal-sheet-scanning OPT setup \cite{Marelli:21}.
To fully characterize the mechanical errors, we adopt the 3D X-ray transform $\mathcal{P}_{\bth}$ that provides a mathematical description of the straight-ray projections of a sample at any 3D pose \cite{Zehni2020} as
%--
\begin{equation}\label{eq:our-continuous-forward}
    b^{\bth,\,\bst}(\bsv)=\mathcal{P}_{\bth}\{f\}(\bsv-\bst) + n,
\end{equation}
%--
where the compactly supported function $f(\bsu)\in L_2(\R^3)$, $\bsu=(x,y,z)$ represents the 3D sample to be reconstructed. 
The measurement in the detector plane is $b^{\bth,\,\bst}(\bsv)$ for a location $\bsv=(\xi,\eta)$, a given sample orientation $\bth$, and a shift vector $\bst$.
The additive random noise is $n$.
To numerically implement the forward model, we discretize Eq. \eqref{eq:our-continuous-forward} under a sampling framework described in \cite{Unser2000} and obtain the following linear system
%-
\begin{equation}\label{eq:final-discrete-forward}
    \mathbf{b}=\mathbf{H}(\boldsymbol{\theta}, \boldsymbol{t})\mathbf{c}+\mathbf{n},
\end{equation}
%-
where $\mathbf{b, \;n}\in\R^{MP}$ are the measurement and noise vectors of $P$ projections and where the system matrix $\mathbf{H}(\boldsymbol{\theta}, \bst)\in\R^{MP\times N}$ is a function of system parameters $\boldsymbol{\theta}$ and $\bst$.
The 3D volume is represented by a finite-dimensional coefficient vector $\mathbf{c}\in\mathbb{R}^N$ using the optimized Kaiser-Bessel window functions that are well suited for tomographic settings \cite{nilchian2015}. This choice of basis is convenient to compute analytic gradients for the tilt angles and shifts \cite{Zehni2020}. 

\subsection{Multiscale Calibration-Reconstruction Algorithm}
\begin{figure}[t]
    \centering
    \includegraphics[width=\linewidth]{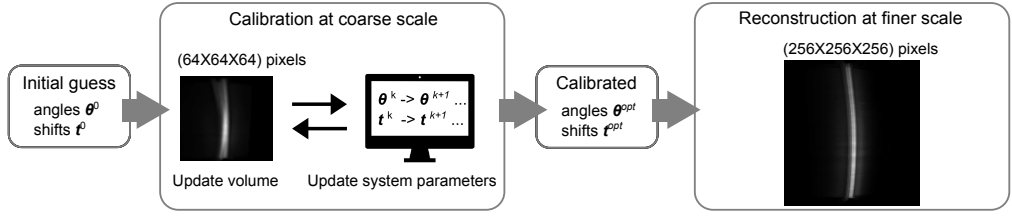}
    \caption{Workflow of the multiscale calibration-reconstruction algorithm. We first perform a joint reconstruction-calibration at coarse scale. We then use the calibrated mechanical parameters for the final high-resolution reconstruction. }
    \label{fig:flow-chart}
\end{figure}

The measurement data generated by a real OPT experiment usually has a very large size. 
For instance, current cameras can acquire OPT images up to size $2048^2$ pixels with pixel size less than 1$\mu$m per projection \cite{Schmidt:21}.
Moreover, the rotation motor can achieve a step angle as precise as 0.3 degrees, resulting in 1200 projections. 
This means the storage of the vector $\mathbf{b}$ can take up to several tens of gigabytes.
3D reconstruction and calibration on such large datasets are either infeasible  memory-wise on GPU or infeasible speed-wise on CPU. 
We thus propose the multiscale scheme illustrated in Fig. \ref{fig:flow-chart} to overcome the computational bottleneck.
We first downsample the original measurement data to a computationally feasible scale and then run our automatic calibration algorithm at coarse scale. 
%--
\begin{algorithm}[t]
    \caption{Automatic Calibration}
    \label{algo:global-recon}
        \begin{algorithmic}[1]
            \REQUIRE $\boldsymbol{\theta}^0,\boldsymbol{t}^0,\mathbf{b},\lambda>0$
            \STATE $\boldsymbol{\theta}=\boldsymbol{\theta}^0, \boldsymbol{t}=\boldsymbol{t}^0$
            \WHILE{$\boldsymbol{\theta}$ and $\boldsymbol{t}$ not converged} 
                \STATE $\mathbf{c}\leftarrow\arg\underset{\mathbf{c}\in\R^N}{\min}\left\{\|\mathbf{H}(\boldsymbol{\theta},\boldsymbol{t})\mathbf{c}-\mathbf{b}\|^2_2\right\}$
                \STATE $(\boldsymbol{\theta},\boldsymbol{t})\leftarrow\arg\underset{\boldsymbol{\theta}\in\R^{P+2},\,\boldsymbol{t}\in\R^{2}}{\min}\left\{\|\mathbf{H}(\boldsymbol{\theta},\boldsymbol{t})\mathbf{c}-\mathbf{b}\|^2_2\right\}$
            \ENDWHILE    
        \RETURN $\boldsymbol{\theta}$ and $\boldsymbol{t}$
        \end{algorithmic}
\end{algorithm}
%--
The inverse problem here is formulated as: 
%--
\begin{equation}\label{eq:global-minimization}
    \mathbf{c}^*,(\boldsymbol{\theta}^*,\boldsymbol{t}^*)\in\arg\underset{\mathbf{c},\boldsymbol{\theta},\boldsymbol{t}}{\min}\left\{\frac{1}{2}\|\mathbf{H}(\boldsymbol{\theta},\boldsymbol{t})\mathbf{c}-\mathbf{b}\|^2_2\right\}.
\end{equation}
%--
The reconstruction pipeline is detailed in Algorithm \ref{algo:global-recon}. 
It consists of the recovery of the coefficient vector~$\mathbf{c}$ by solving the optimization problem in Step 3 and calibrating the system parameters $\boldsymbol{\theta}$ and $\boldsymbol{t}$ by solving another optimization problem in Step 4 in alternating fashion.
We start with an initial guess for the system parameter $\boldsymbol{\theta}^0$ and $\boldsymbol{t}^0=(0, 0)$ that corresponds to a perfect setup.
Inspired by the observations of Section 2.3, which indicate that the radius of the circle artifacts is directly related to value of the shift, we shift the projections over a range of values and then apply the FBP algorithm to observe the evolution of the circle artifacts in the reconstruction.
This allows us to attain a relatively close initial guess for $\boldsymbol{t}$ in a fast manner, which is crucial for the convergence of the calibration algorithm. 
Such a set of $\boldsymbol{\theta}$ and $\boldsymbol{t}$ serves as a good initial guess for the calibration algorithm.

The alternating process is repeated until the system parameters are well refined. 
The calibrated system parameters are then used to reconstruct an artifact-free 3D image by running an extra Step 3 in Algorithm \ref{algo:global-recon} at a finer scale, as described in Fig. \ref{fig:flow-chart}.

\subsection{Results on Simulated and Experimental Data}
To validate Algorithm \ref{algo:global-recon}, we simulate the ground truth as a 3D cube of size $512^3$ pixels in which we have randomly inserted 150 beads.
A positive constant accumulative error of 0.05 degrees in the rotation angles and a constant shift of 8 pixels along the $x$ axis are added to the forward model to simulate a set of 300 projections of realistic OPT measurements. 
Then, we downsampled the measurements to a coarse scale of ($128\times128\times300$) pixels and applied Algorithm \ref{algo:global-recon}. 
The initial guess was set as described in Section 3.2. 
The reconstruction step of the algorithm is implemented using the GlobalBioIm library \cite{Soubies:19} and the calibration step uses functions from the Cryo-refinement library \cite{Zehni2020}.

\begin{figure}[t]
    \centering
    \includegraphics[width=\textwidth]{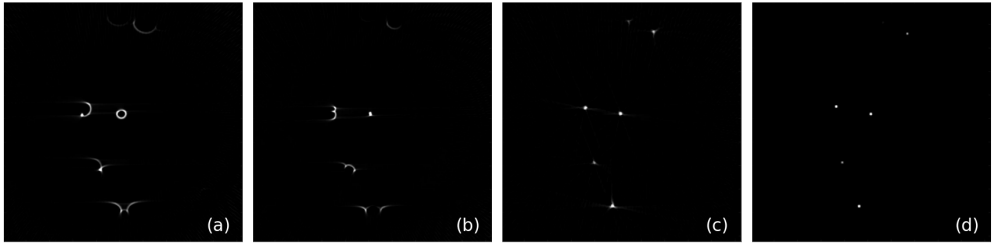}
    \caption{(a) Reconstruction with circle and seagull artifacts at one slice (147) without calibration. (b) Reconstruction result at slice 147 after a naive shift correction. (c) Calibrated reconstruction at slice 147. Both the circle and seagull artifacts are successfully removed. (d) Ground truth at slice 147. Images are saturated for better visualization.}
    \label{fig:simu_calibration}
\end{figure}

In total, 10 global rounds of joint reconstruction-calibration were used, each of which composed of only 30 iterations of reconstruction and 6 iterations of calibration to avoid overfitting in the presence of model mismatch. 
After having obtained the calibrated system parameters, we reconstructed the 3D volume at a finer scale ($512^3$) using the FBP algorithm.
In Fig. \ref{fig:simu_calibration}(a), we show that the reconstruction without any calibration or shift correction contains both the circle and seagull-shaped artifacts.
The reconstruction with only naive shift correction (Fig. \ref{fig:simu_calibration}(b)) still suffers from seagull artifacts due to residual model mismatch.
After applying our calibration algorithm, both the circle and the seagull artifacts are successfully removed all at once (Fig. \ref{fig:simu_calibration}(c)) and the reconstruction is very close to the ground-truth image in Fig.  \ref{fig:simu_calibration}(d).
%===========================================================

Despite the best effort of the experimentalists to calibrate the hardware system, the rotation axis may still contain undesirable tilting which degrades the quality of the reconstructed OPT images.
Similar to the procedures in the simulation, we follow the steps in the flowchart in Fig. \ref{fig:flow-chart} to further validate our algorithm on an experimental dataset of a fluorescent textile fiber that contains errors in the the tilt angles.
The data are acquired using a focal-sheet-scanning OPT system \cite{Marelli:21}.
It uses a lateral light-sheet illumination to reduce the out-of-focus blur in the images, enabling the assumption of straight-ray projections for our forward model. We use $720$ projections of size ($256 \times 256$) that we downsample to a coarse scale of ($64 \times 64 \times 720$) pixels.

\begin{figure}[t]
    \centering
    \includegraphics[width=\textwidth]{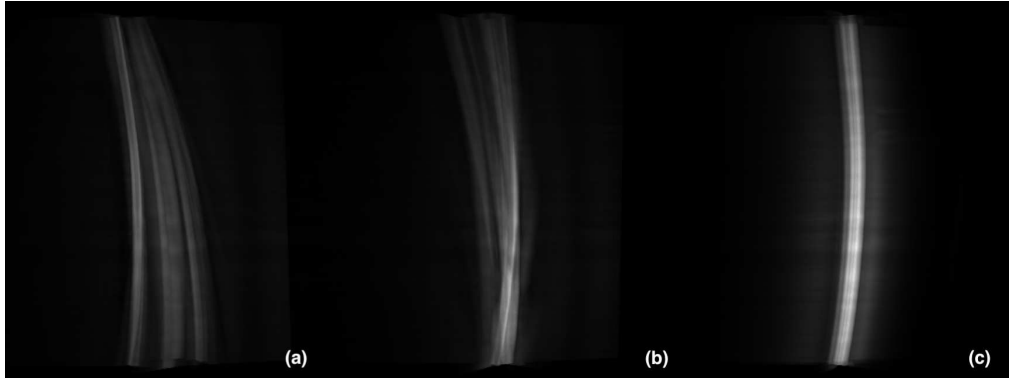}
    \caption{Textile fiber data. (a) - (b) Snapshots of the 3D visualization of the reconstruction results using the 2D FBP algorithm in a slice-by-slice fashion with no correction (a) and correcting only for the center of rotation (b). (c) 3D reconstruction result using the calibrated system parameters. }
    \label{fig:fiber}
\end{figure}

\begin{figure}[t]
    \centering
    \includegraphics[width=0.4\linewidth]{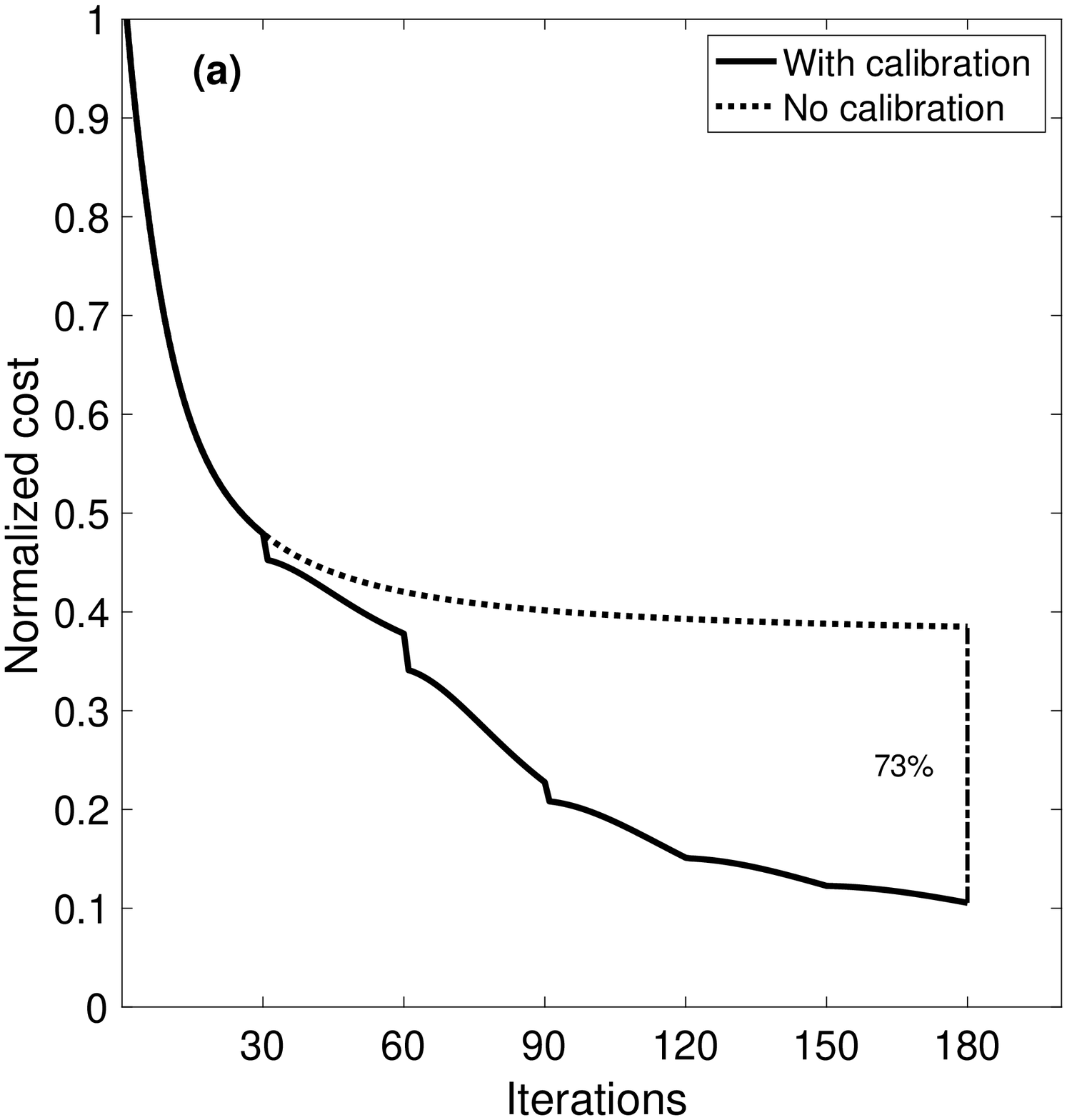}
    \includegraphics[width=0.4\linewidth]{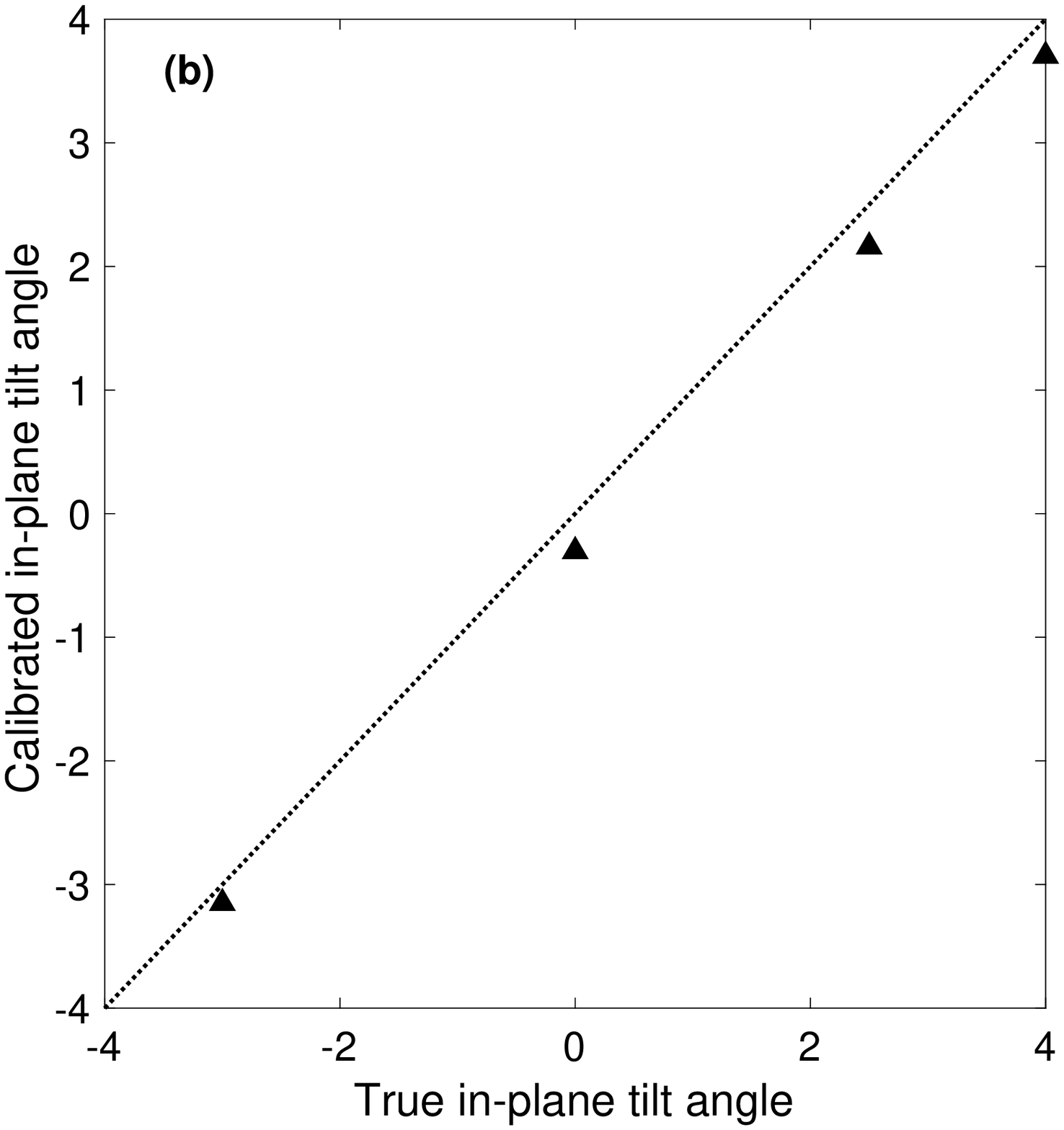}
    \caption{
    Convergence and accuracy. (a) Evolution of the cost function of the reconstruction during the calibration process with (solid) and without (dashed) calibration. (b) The calibrated in-plane tilt angle against the true in-plane tilt angle for 4 different values indicated by the triangles. The dotted line represents $y=x$. The closer the triangles are to this line, the closer the calibrated angles are to the true angles.}
    \label{fig:stats}
\end{figure}

We show the four tilt angles after calibration in Fig. \ref{fig:stats}(b). 
All four scenarios of different magnitudes and directions of tilt angles led to success as the calibrated angle is very close to the controlled true value.
The 3D visualization of the reconstruction result of the fiber with an out-of plane tilt angle of approximately 4 degrees is displayed in Fig. \ref{fig:fiber}. 
Without any correction, the reconstruction shows multiple ghost artifacts due to a combination of misaligned COR and tilt of the rotation axis in Fig. \ref{fig:fiber}(a). 
The severity of these ghost shadows are reduced after naive shift correction, as seen in Fig. \ref{fig:fiber}(b) but the top and middle parts of the reconstructed fiber still suffer from ghost shadows.
The 3D reconstruction using the calibrated system parameters output by Algorithm \ref{algo:global-recon} effectively removes all the shadows and achieves an artifact-free image.
The evolution of the cost function during the joint reconstruction-calibration is displayed in Fig. \ref{fig:stats}(a) and shows a 73\% reduction of the cost compared to the uncalibrated configuration. 
After 6 global rounds, the calibration algorithm found an out-of-plane tilt angle of 3.7 degrees which is very close to the controlled 4 degrees.
This further confirms that the calibration manages to reduce the model mismatch due to tilt. 
%--

\section{Conclusion}
We presented a comprehensive study of certain geometrical artifacts in a poorly calibrated OPT imaging system.
We summarized our results in the form of a catalog by combining existing results on the shifted center of rotation and our own contribution on the 3D tilt of the rotation axis and the inaccurate rotation angles.
In doing so, we are able to explain the various types of mechanical artifacts, its appearance, cause, and properties, as well as the associated correction methods.
This catalog serves as a reference for OPT practitioners to gain insight into their experimental setup and help them better calibrate their hardware system.
To fill the vacancy of a versatile computational method to account for all types of mechanical artifacts, we propose an automatic calibration algorithm.
It is based on a refined 3D mathematical model of the OPT imaging geometry that is able to characterize mechanical errors in the system. 
Moreover, the algorithm adapts to the large-size measurement datasets of OPT by performing the calibration on a coarse scale to overcome the computational bottleneck, while the final reconstruction of the 3D volume is achieved on a finer scale. 
Our multiscale calibration scheme has first been validated on the synthesized bead data where we simulate the shifted center of rotation and the imprecise rotation angles then successfully remove the resulting artifacts.
We have further applied our algorithm on an experimental dataset of a fluorescent textile fiber that suffers from a tilted rotation axis, managing to detect the model mismatch and recover the true tilt angle.
In the visualization of the final reconstructed volume, the associated artifacts are taken care of and we obtain a clean 3D image.

\begin{backmatter}
\bmsection{Funding}
Y. L., J.D., T.A.P., and M.U. acknowledge funding from European Research Council (ERC) under the European Union’s Horizon 2020 research and innovation programme (Grant Agreement No. 692726 GlobalBioIm). F.M. acknowledges funding from the Swiss NSF under grant numbers 200020\_179217: Computational biomicroscopy: Advanced Image Processing Methods to Quantify Live Biological Systems, and 206021\_164022: Platform for Reproducible Acquisition, Processing, and Sharing of Dynamic, Multi-Modal Data.

\bmsection{Disclosures}
The authors declare no conflicts of interest.

\bmsection{Data availability} 
Data and code underlying the textile-fiber results may be available upon request.
\end{backmatter}

\bibliography{ref.bib}

\begin{thebibliography}{10}
\newcommand{\enquote}[1]{``#1''}

\bibitem{Sharpe2004}
J.~Sharpe, \enquote{Optical projection tomography,}
  {\protect\JournalTitle{Annual Review of Biomedical Engineering}} \textbf{6},
  209--228 (2004).

\bibitem{Lindsey2017}
B.~W. Lindsey and J.~Kaslin, \enquote{Optical projection tomography as a novel
  method to visualize and quantitate whole-brain patterns of cell proliferation
  in the adult zebrafish brain,} {\protect\JournalTitle{Zebrafish}}
  \textbf{14}, 574--577 (2017).

\bibitem{Schmidt:21}
C.~Schmidt, A.~L. Planchette, D.~Nguyen, G.~Giardina, Y.~Neuenschwander, M.~D.
  Franco, A.~Mylonas, A.~C. Descloux, E.~Pomarico, A.~Radenovic, and
  J.~Extermann, \enquote{High resolution optical projection tomography platform
  for multispectral imaging of the mouse gut,} {\protect\JournalTitle{Biomed.
  Opt. Express}} \textbf{12}, 3619--3629 (2021).

\bibitem{Wang:07}
Y.~Wang and R.~K. Wang, \enquote{Optimization of image-forming optics for
  transmission optical projection tomography,} {\protect\JournalTitle{Appl.
  Opt.}} \textbf{46}, 6815--6820 (2007).

\bibitem{bassi:11}
A.~Bassi, L.~Fieramonti, C.~D'Andrea, G.~Valentini, and M.~Mione, \enquote{{In
  vivo label-free three-dimensional imaging of zebrafish vasculature with
  optical projection tomography},} {\protect\JournalTitle{Journal of Biomedical
  Optics}} \textbf{16}, 1--4 (2011).

\bibitem{Walls2007}
J.~R. Walls, J.~G. Sled, J.~Sharpe, and R.~M. Henkelman, \enquote{Resolution
  improvement in emission optical projection tomography,}
  {\protect\JournalTitle{Physics in Medicine and Biology}} \textbf{52},
  2775--2790 (2007).

\bibitem{Walls2005}
J.~R. Walls, J.~G. Sled, J.~Sharpe, and R.~M. Henkelman, \enquote{Correction of
  artefacts in optical projection tomography,} {\protect\JournalTitle{Physics
  in Medicine and Biology}} \textbf{50}, 4645--4665 (2005).

\bibitem{McGinty:11}
J.~McGinty, H.~B. Taylor, L.~Chen, L.~Bugeon, J.~R. Lamb, M.~J. Dallman, and
  P.~M.~W. French, \enquote{In vivo fluorescence lifetime optical projection
  tomography,} {\protect\JournalTitle{Biomed. Opt. Express}} \textbf{2},
  1340--1350 (2011).

\bibitem{Torres2021}
V.~C. Torres, C.~Li, W.~Zhou, J.~G. Brankov, and K.~M. Tichauer,
  \enquote{Characterization of an angular domain fluorescence optical
  projection tomography system for mesoscopic lymph node imaging,}
  {\protect\JournalTitle{Appl. Opt.}} \textbf{60}, 135--146 (2021).

\bibitem{Sharpe2002}
J.~Sharpe, U.~Ahlgren, P.~Perry, B.~Hill, A.~Ross, J.~Hecksher-Sørensen,
  R.~Baldock, and D.~Davidson, \enquote{Optical projection tomography as a tool
  for 3{D} microscopy and gene expression studies,}
  {\protect\JournalTitle{Science}} \textbf{296}, 541--545 (2002).

\bibitem{Natterer}
F.~Natterer, \emph{The Mathematics of Computerized Tomography} (Society for
  Industrial and Applied Mathematics, USA, 2001).

\bibitem{Vinegoni:09}
C.~Vinegoni, L.~Fexon, P.~F. Feruglio, M.~Pivovarov, J.-L. Figueiredo,
  M.~Nahrendorf, A.~Pozzo, A.~Sbarbati, and R.~Weissleder, \enquote{High
  throughput transmission optical projection tomography using low cost graphics
  processing unit,} {\protect\JournalTitle{Opt. Express}} \textbf{17},
  22320--22332 (2009).

\bibitem{Ramirez:19}
P.~P. Vallejo~Ramirez, J.~Zammit, O.~Vanderpoorten, F.~Riche, F.-X. Bl{\'e},
  X.-H. Zhou, B.~Spiridon, C.~Valentine, S.~E. Spasov, P.~W. Oluwasanya,
  G.~Goodfellow, M.~J. Fantham, O.~Siddiqui, F.~Alimagham, M.~Robbins,
  A.~Stretton, D.~Simatos, O.~Hadeler, E.~J. Rees, F.~Str{\"o}hl, R.~F. Laine,
  and C.~F. Kaminski, \enquote{{OptiJ}: Open-source optical projection
  tomography of large organ samples,} {\protect\JournalTitle{Scientific
  Reports}} \textbf{9}, 15693 (2019).

\bibitem{Zhang:20}
H.~Zhang, L.~Waldmann, R.~Manuel, H.~Boije, T.~Haitina, and A.~Allalou,
  \enquote{{zOPT}: an open source optical projection tomography system and
  methods for rapid 3{D} zebrafish imaging,} {\protect\JournalTitle{Biomed.
  Opt. Express}} \textbf{11}, 4290--4305 (2020).

\bibitem{Tang:16}
X.~Tang, M.~van't Hoff, J.~Hoogenboom, Y.~Guo, F.~Cai, G.~Lamers, and
  F.~Verbeek, \enquote{Fluorescence and bright-field {3D} image fusion based on
  sinogram unification for optical projection tomography,} in \emph{2016 IEEE
  International Conference on Bioinformatics and Biomedicine (BIBM),}  (2016),
  pp. 403--410.

\bibitem{Donath:06}
T.~Donath, F.~Beckmann, and A.~Schreyer, \enquote{Automated determination of
  the center of rotation in tomography data,} {\protect\JournalTitle{J. Opt.
  Soc. Am. A}} \textbf{23}, 1048--1057 (2006).

\bibitem{michalek_2015}
J.~Michalek, \enquote{Total variation-based reduction of streak artifacts, ring
  artifacts and noise in {3D} reconstruction from optical projection
  tomography,} {\protect\JournalTitle{Microscopy and Microanalysis}}
  \textbf{21}, 1602–1615 (2015).

\bibitem{Koskela19}
O.~Koskela, T.~Montonen, B.~Belay, E.~Figueiras, S.~Pursiainen, and
  J.~Hyttinen, \enquote{Gaussian light model in brightfield optical projection
  tomography,} {\protect\JournalTitle{Scientific Reports}} \textbf{9} (2019).

\bibitem{Koljonen19}
V.~Koljonen, O.~Koskela, T.~Montonen, A.~Rezaei, B.~Belay, E.~Figueiras,
  J.~Hyttinen, and S.~Pursiainen, \enquote{A mathematical model and iterative
  inversion for fluorescent optical projection tomography,}
  {\protect\JournalTitle{Physics in Medicine \& Biology}} \textbf{64}, 045017
  (2019).

\bibitem{Trull2017}
A.~K. Trull, J.~van~der Horst, W.~J. Palenstijn, L.~J. van Vliet, T.~van
  Leeuwen, and J.~Kalkman, \enquote{Point spread function based image
  reconstruction in optical projection tomography,}
  {\protect\JournalTitle{Physics in Medicine \& Biology}} \textbf{62},
  7784--7797 (2017).

\bibitem{Birk2011}
U.~J. Birk, A.~Darrell, N.~Konstantinides, A.~Sarasa-Renedo, and J.~Ripoll,
  \enquote{Improved reconstructions and generalized filtered back projection
  for optical projection tomography,} {\protect\JournalTitle{Appl. Opt.}}
  \textbf{50}, 392--398 (2011).

\bibitem{Antonopoulos:14}
G.~C. Antonopoulos, D.~Pscheniza, R.-A. Lorbeer, M.~Heidrich, K.~Schwanke,
  R.~Zweigerdt, T.~Ripken, and H.~Meyer, \enquote{{Correction of image
  artifacts caused by refractive index gradients in scanning laser optical
  tomography},} in \emph{Three-Dimensional and Multidimensional Microscopy:
  Image Acquisition and Processing XXI,}  vol. 8949 (2014), p. 894907.

\bibitem{liu2022artifacts}
Y.~Liu, J.~Dong, C.~Schmidt, A.~Boquet-Pujadas, J.~Extermann, and M.~Unser,
  \enquote{Artifacts in optical projection tomography due to refractive-index
  mismatch: {Model} and correction,} {\protect\JournalTitle{Optics Letters}}
  \textbf{47}, 2618--2621 (2022).

\bibitem{vanderHorst:16}
J.~van~der Horst and J.~Kalkman, \enquote{Image resolution and deconvolution in
  optical tomography,} {\protect\JournalTitle{Opt. Express}} \textbf{24},
  24460--24472 (2016).

\bibitem{Darrell2008}
A.~Darrell, H.~Meyer, K.~Marias, M.~Brady, and J.~Ripoll, \enquote{Weighted
  filtered backprojection for quantitative fluorescence optical projection
  tomography,} {\protect\JournalTitle{Physics in Medicine \& Biology}}
  \textbf{53}, 3863--3881 (2008).

\bibitem{Darrell_MLE}
A.~Darrell, H.~Meyer, U.~Birk, K.~Marias, M.~Brady, and J.~Ripoll,
  \enquote{Maximum likelihood reconstruction for fluorescence optical
  projection tomography,} in \emph{2008 8th IEEE International Conference on
  BioInformatics and BioEngineering,}  (2008), pp. 1--6.

\bibitem{CT2001}
A.~C. Kak and M.~Slaney, \emph{Principles of Computerized Tomographic Imaging}
  (Society for Industrial and Applied Mathematics, 2001).

\bibitem{Birk:10}
U.~J. Birk, M.~Rieckher, N.~Konstantinides, A.~Darrell, A.~Sarasa-Renedo,
  H.~Meyer, N.~Tavernarakis, and J.~Ripoll, \enquote{Correction for specimen
  movement and rotation errors for in-vivo optical projection tomography,}
  {\protect\JournalTitle{Biomed. Opt. Express}} \textbf{1}, 87--96 (2010).

\bibitem{Ancora2017}
D.~Ancora, D.~D. Battista, G.~Giasafaki, S.~Psycharakis, E.~Liapis,
  A.~Zacharopoulos, and G.~Zacharakis, \enquote{{Optical projection tomography
  via phase retrieval algorithms for hidden three dimensional imaging},} in
  \emph{Quantitative Phase Imaging III,}  vol. 10074 (SPIE, 2017), p. 100741E.

\bibitem{Hendriksen:21}
A.~A. Hendriksen, D.~Schut, W.~J. Palenstijn, N.~Vigan\'{o}, J.~Kim, D.~M.
  Pelt, T.~van Leeuwen, and K.~J. Batenburg, \enquote{Tomosipo: fast, flexible,
  and convenient 3d tomography for complex scanning geometries in python,}
  {\protect\JournalTitle{Opt. Express}} \textbf{29}, 40494--40513 (2021).

\bibitem{Marelli:21}
F.~Marelli and M.~Liebling, \enquote{Optics versus computation: Influence of
  illumination and reconstruction model accuracy in focal-plane-scanning
  optical projection tomography,} in \emph{2021 IEEE 18th International
  Symposium on Biomedical Imaging (ISBI),}  (2021), pp. 567--570.

\bibitem{Zehni2020}
M.~Zehni, L.~Donati, E.~Soubies, Z.~Zhao, and M.~Unser, \enquote{Joint angular
  refinement and reconstruction for single-particle cryo-em,}
  {\protect\JournalTitle{IEEE Transactions on Image Processing}} \textbf{29},
  6151--6163 (2020).

\bibitem{Unser2000}
M.~Unser, \enquote{Sampling---50 {Years} after {Shannon},}
  {\protect\JournalTitle{Proceedings of the IEEE}} \textbf{88}, 569--587
  (2000).

\bibitem{nilchian2015}
M.~Nilchian, J.~P. Ward, C.~Vonesch, and M.~Unser, \enquote{Optimized
  {Kaiser–Bessel} window functions for computed tomography,}
  {\protect\JournalTitle{IEEE Transactions on Image Processing}} \textbf{24},
  3826--3833 (2015).

\bibitem{Soubies:19}
E.~Soubies, F.~Soulez, M.~T. McCann, {\exhyphenpenalty9999\relax{}T.-a.}~Pham,
  L.~Donati, T.~Debarre, D.~Sage, and M.~Unser, \enquote{Pocket guide to solve
  inverse problems with {GlobalBioIm},} {\protect\JournalTitle{Inverse
  Problems}} \textbf{35}, 104006 (2019).

\end{thebibliography}
\clearpage
\end{document}